\documentclass[12pt,a4paper]{iopart}
\usepackage{setstack,iopams}
\usepackage{graphicx} 
\usepackage[caption=false]{subfig}
\begin{document}


\newcommand{\braket}[2]{{\left\langle #1 \middle| #2 \right\rangle}}
\newcommand{\bra}[1]{{\left\langle #1 \right|}}
\newcommand{\ket}[1]{{\left| #1 \right\rangle}}
\newcommand{\ketbra}[2]{{\left| #1 \middle\rangle \middle \langle #2 \right|}}


\title{Quantum Walk Search with Time-Reversal Symmetry Breaking}

\author{Thomas G Wong}
\address{Faculty of Computing, University of Latvia, Rai\c{n}a bulv.~19, R\=\i ga, LV-1586, Latvia}
\ead{\mailto{twong@lu.lv}}

\begin{abstract}
	We formulate Grover's unstructured search algorithm as a chiral quantum walk, where transitioning in one direction has a phase conjugate to transitioning in the opposite direction. For small phases, this breaking of time-reversal symmetry is too small to significantly affect the evolution: the system still approximately evolves in its ground and first excited states, rotating to the marked vertex in time $\pi \sqrt{N} / 2$. Increasing the phase does not change the runtime, but rather changes the support for the 2D subspace, so the system evolves in its first and second excited states, or its second and third excited states, and so forth. Apart from the critical phases corresponding to these transitions in the support, which become more frequent as the phase grows, this reveals that our model of quantum search is robust against time-reversal symmetry breaking.
\end{abstract}

\pacs{03.67.Ac}


\section{Introduction}

Continuous-time quantum walks \cite{FG1998b,Kempe2003} are analogues of continuous-time classical Markov chains. In them, probability amplitude flows from one vertex of a graph to another by evolving by Schr\"odinger's equation
\[ i \frac{\rmd}{\rmd t} \ket{\psi} = H \ket{\psi} \]
with Hamiltonian
\begin{equation}
	\label{eq:H_walk}
	H_{\rm walk} = -\gamma L,
\end{equation}
where $L = A-D$ is the graph Laplacian, which is composed of the adjacency matrix $A$ for which $A_{ij} = 1$ when vertices $i$ and $j$ are connected (and zero otherwise) and the diagonal degree matrix $D$ for which $D_{ii} = {\rm deg}(i)$. The real parameter $\gamma$ is the jumping rate, or amplitude per time. This evolution causes probability amplitude to transition between adjacent vertices.

\begin{figure}
\begin{center}
	\subfloat[] {
		\includegraphics{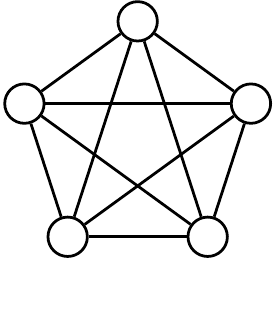}
		\label{fig:complete}
	} \quad
	\subfloat[] {
		\includegraphics[height=1.5in]{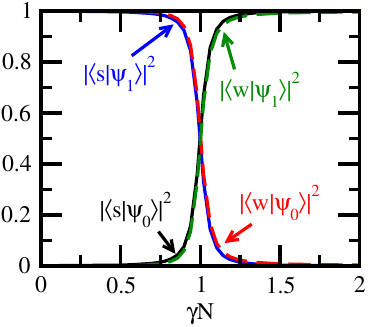}
		\label{fig:overlap_N1023_theta0}
	} \quad
	\subfloat[] {
		\includegraphics[height=1.5in]{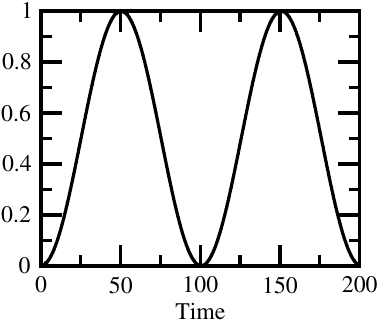}
		\label{fig:prob_time_N1023_theta0}
	}
	\caption{(a) The complete graph of $N = 5$ vertices. (b) Probability overlaps of $\ket{s}$ and $\ket{w}$ with the ground and first excited states $\ket{\psi_{0,1}}$ of $H$ for quantum walk search with $N = 1023$. (c) Success probability for quantum walk search with $N = 1023$ and $\gamma = \gamma_c = 1/N$.}
\end{center}
\end{figure}

For example, consider the quantum walk on the complete graph of $N$ vertices, an example of which is shown in \fref{fig:complete}. Then the vertices label computational basis states $\{ \ket{1}, \dots, \ket{N} \}$ of an $N$-dimensional Hilbert space. Since each vertex is connected to the $N-1$ others, the Laplacian is an $N \times N$ matrix with $-(N-1)$ on the diagonal and $1$ everywhere else:
\[ L = \left( \begin{array}{cccc}
	-(N-1) & 1 & \cdots & 1 \\
	1 & -(N-1) & \cdots & 1 \\
	\vdots & \vdots & \ddots & \vdots \\
	1 & 1 & \cdots & -(N-1) \\
\end{array} \right). \]
If the system starts in the equal superposition over all the vertices
\[ \ket{s} = \frac{1}{\sqrt{N}} \sum_{i = 1}^N \ket{i}, \]
then the system stays in this equilibrium state, up to a global phase, when evolving by the quantum walk.

To construct a search algorithm from the quantum walk, we include an additional term in the Hamiltonian
\begin{equation}
	\label{eq:H}
	H = -\gamma L - \ketbra{w}{w},
\end{equation}
which marks a basis state $\ket{w}$ and acts as an oracle \cite{Mochon2007}. The goal is to find $\ket{w}$ by evolving with this Hamiltonian, and we begin in $\ket{s}$ to express our initial lack of knowledge of which vertex $\ket{w}$ might be. Note that the complete graph is homogeneous, so which particular vertex is marked does not change the dynamics. Also, each vertex is connected to every other, and there is no additional structure to the problem. Thus this is the quantum walk formulation of the unstructured, black-box search problem that Grover's algorithm \cite{Grover1996} famously solves.

This formulation of Grover's algorithm is quite easy to solve, as shown by Childs and Goldstone \cite{CG2004}, and which we summarize here. Since the unmarked vertices evolve identically, the system evolves in a 2D subspace. As shown in \fref{fig:overlap_N1023_theta0}, when $\gamma$ takes its critical value of $\gamma_c = 1/N$, this 2D subspace is spanned by eigenstates of $H$
\[ \ket{\psi_{0,1}} \propto \ket{s} \mp \ket{w}, \]
which have an energy gap of $\mathrm{\Delta} E = E_1 - E_0 = 2/\sqrt{N}$. Thus the system evolves from $\ket{s}$ to $\ket{w}$ with probability $1$ in time $\pi/\mathrm{\Delta} E = \pi \sqrt{N} / 2$, and a detailed calculation showing this is available in Section III of \cite{Wong2015b}. The evolution of the success probability with time is shown in \fref{fig:prob_time_N1023_theta0}, and it reaches $1$ at $\pi \sqrt{1023} / 2 \approx 50.241$, as expected.

\begin{figure}
\begin{center}
	\subfloat[] {
		\includegraphics{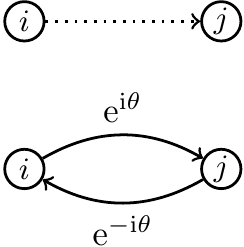}
		\label{fig:notation}
	} \quad \quad
	\subfloat[] {
		\includegraphics{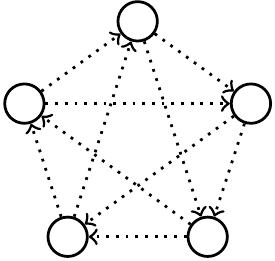}
		\label{fig:complete_chiral}
	}
	\caption{(a) Graphic notation for chiral quantum walks. A dotted arrow from $i$ to $j$ denotes a directed edge from $i$ to $j$ with weight $\rme^{\rmi\theta}$, and from $j$ to $i$ with weight $\rme^{-\rmi\theta}$. (b) Chiral quantum walk on the complete graph with $N = 5$.}
\end{center}
\end{figure}

Such quantum walks are the basis of several of quantum algorithms \cite{SKW2003,FGG2008} and even universal computation \cite{Childs2009}, and they preserve time-reversal symmetry, meaning the probability at each vertex at time $t$ is the same as at time $-t$. Recently, Zimbor\'as \textit{et al.} introduced \emph{chiral quantum walks} \cite{ZFKWLB2013}, which break time-reversal symmetry by walking on directed, weighted graphs, where transitioning in one direction has some phase $\rme^{\rmi\theta}$, and transitioning in the opposite direction has the conjugate phase $\rme^{-\rmi\theta}$. We introduce a graphic notation for this in \fref{fig:notation}, where a dotted arrow from vertex $i$ to $j$ indicates a directed edge from $i$ to $j$ with weight $\rme^{\rmi\theta}$, and from $j$ to $i$ with weight $\rme^{-\rmi\theta}$.

By breaking time-reversal symmetry, Zimbor\'as \textit{et al.} attained significant speedups in quantum transport, indicating that such breaking may be a useful tool in engineering hardware to efficiently transport energy or information. While faster transport with a directed quantum walk \cite{HM2009} is unsurprising since allowing transitions in just one direction provides immense control over the flow of amplitude, speedups with chiral quantum walks are novel because the conjugate relationship that opposite directions must maintain prevents one from eliminating one direction of the walk entirely. Chiral quantum walks have also been investigated in discrete-time \cite{LBLLJBFZLBL2014}, but here we focus on the continuous-time formulation.

Given the speedups obtained by Zimbor\'as \textit{et al.} in quantum transport, one might hope to achieve similar speedups algorithmically. In this paper, we conduct the first \emph{algorithmic} investigation of chiral quantum walks, showing how breaking time-reversal symmetry affects quantum search. While the $O(\sqrt{N})$ optimality \cite{FG1998a} of the continuous-time search Hamiltonian \eref{eq:H} prevents time-reversal symmetry breaking from improving the runtime scaling, it is possible that it improves the constant factor. Of course, it is also possible that it makes the search slower. We show that neither of these occur, that the runtime is unchanged from $\pi \sqrt{N}/2$ for large swathes of $\theta$. In spite of this, the value of $\gamma$ must be chosen differently, and the system may not evolve in its lowest two energy eigenstates.

In particular, we modify the complete graph as shown in \fref{fig:complete_chiral}. Assuming throughout the paper that $N$ is odd, we arrange the vertices in a circular pattern. Then using our graphic notation in \fref{fig:notation} for chiral quantum walks, each vertex has a dotted arrow going from it to the $(N-1)/2$ vertices ahead (in a clockwise direction) of it, and the $(N-1)/2$ vertices behind it have dotted arrows coming into the vertex. Thus each vertex has half its dotted arrows going out and half coming in, but in such a way that the graph is homogeneous, so it makes no difference as to which vertex is marked for search.

To further illuminate the chiral quantum walk we have defined on the modified complete graph in \fref{fig:complete_chiral}, the graph Laplacian $L$ for it when $N = 5$ is
\[ L = \left( \begin{array}{ccccc}
	-N_1\cos\theta & \rme^{-\rmi\theta} & \rme^{-\rmi\theta} & \rme^{\rmi\theta} & \rme^{\rmi\theta} \\
	\rme^{\rmi\theta} & -N_1\cos\theta & \rme^{-\rmi\theta} & \rme^{-\rmi\theta} & \rme^{\rmi\theta} \\
	\rme^{\rmi\theta} & \rme^{\rmi\theta} & -N_1\cos\theta & \rme^{-\rmi\theta} & \rme^{-\rmi\theta} \\
	\rme^{-\rmi\theta} & \rme^{\rmi\theta} & \rme^{\rmi\theta} & -N_1\cos\theta & \rme^{-\rmi\theta} \\
	\rme^{-\rmi\theta} & \rme^{-\rmi\theta} & \rme^{\rmi\theta} & \rme^{\rmi\theta} & -N_1\cos\theta \\
\end{array} \right), \]
where $N_1 = N-1$. Generalizing this to arbitrary $N$ is straightforward---the first column starts with $-(N-1)\cos\theta$, is followed by $(N-1)/2$ terms that are $\rme^{\rmi\theta}$, and ends with $(N-1)/2$ terms that are $\rme^{-\rmi\theta}$. The remaining columns are given by cyclic permutations of this first column, so $L$ is a circulant matrix \cite{Davis1994}. Note that the equal superposition $\ket{s}$ is still the equilibrium distribution of this chiral quantum walk \eref{eq:H_walk}, as in the case of the normal ($\theta = 0$) quantum walk.

The symmetry of this graph Laplacian $L$ restricts the values of $\theta$ we must consider. For example, when $\theta > \pi$, then $\rme^{\pm\rmi \theta} = \rme^{\mp\rmi(\theta - \pi)}$, so we can simply use $\theta - \pi \le \pi$ and reverse all the arrows in \fref{fig:complete_chiral}, which is an equivalent definition of the walk. Thus we can restrict $\theta$ to the interval $[0, \pi]$. Furthermore, when $\theta > \pi/2$, then $\rme^{\pm\rmi \theta} = -\rme^{\mp\rmi(\pi - \theta)}$, so we can instead use $\pi - \theta < \pi/2$, reverse the arrows in \fref{fig:complete_chiral}, and flip the sign of $\gamma$ in \eref{eq:H_walk}. Thus we can restrict $\theta$ to the interval $[0,\pi/2]$, which we do for the rest of the paper.

To turn this chiral quantum walk into a search algorithm, we again include an additional oracle term in the Hamiltonian \eref{eq:H} that marks the vertex to search for. When $\theta > 0$, the unmarked vertices no longer evolve identically, and so the system no longer evolves in an exact 2D subspace. Thus finding the runtime and success probability of the algorithm by finding the eigenstates and eigenvalues of $H$ is non-trivial. Nonetheless, we solve this problem using the method of Childs and Goldstone \cite{CG2004}, where various sums involving the eigenvalues of the negative Laplacian $-L$ can be used to determine the critical $\gamma$, as well as the runtime and success probability of the algorithm. We summarize this method in the next section. Afterward, we solve search by chiral quantum walk, showing that the system still evolves in an approximate 2D subspace, but the value of $\theta$ can cause the support of the subspace by the eigenvectors of $H$ to change. Away from the critical $\theta$'s at which the support changes, breaking time-reversal symmetry does not affect the search algorithm, so it still succeeds with probability $1$ at time $\pi \sqrt{N} / 2$.


\section{\label{Section:CG}Review of Childs and Goldstone's Method}

We begin by reviewing Childs and Goldstone's \cite{CG2004} method, which they used to solve spatial search on arbitrary-dimensional cubic lattices by quantum walk. We will not rederive it, but simply state how to use it. Recall that the search Hamiltonian \eref{eq:H} has two terms: the quantum walk term $-\gamma L$ \eref{eq:H_walk} and the oracle $-\ketbra{w}{w}$. In general, the eigenvalues and eigenvectors of the search Hamiltonian \eref{eq:H} are hard to find, but often they are easy to find for the walk term alone because of the symmetry of the graph.

In particular, let $\mathcal{E}_j$ be the eigenvalues of $-L$, with $j = 0, 1, \dots, N-1$ labeling the eigenvalues. Then consider the sum from Childs and Goldstone's (33):
\begin{equation}
	\label{eq:CG_sum}
	S_i = \frac{1}{N} \sum_{\mathcal{E}_j \ne 0} \frac{1}{(\mathcal{E}_j)^i}.
\end{equation}
From their analysis in Section IV.B., the critical $\gamma$ corresponds to $S_1$ for large $N$:
\begin{equation}
	\label{eq:CG_gammac}
	\gamma_c \approx S_1.
\end{equation}
As shown in \fref{fig:overlap_N1023_theta0}, when $\gamma < \gamma_c$, the initial equal superposition state $\ket{s}$ is approximately an eigenstate of $H$, and when $\gamma > \gamma_c$, it is approximately another eigenstate of $H$. Thus the overlap of $\ket{s}$ with the eigenstates of $H$ exhibits a phase transition at $\gamma_c$.

As an example, for the normal quantum walk on the complete graph (which is the chiral quantum walk with $\theta = 0$), $-L$ has eigenvalues $0$ with multiplicity $1$ and $N$ with multiplicity $N-1$ \cite{Chung1997}. Then
\[ S_1 = \frac{1}{N} \frac{N-1}{N} \approx \frac{1}{N}, \]
which is the expected $\gamma_c$ from the introduction.

From Childs and Goldstone's (96), the success probability is approximately
\begin{equation}
	\label{eq:CG_prob}
	p_* \approx \frac{S_1}{\sqrt{S_2}}
\end{equation}
at time
\begin{equation}
	\label{eq:CG_runtime}
	t_* \approx \frac{\pi}{2} \frac{\sqrt{S_2}}{S_1} \sqrt{N}.
\end{equation}
Continuing the above example of normal quantum walk search on the complete graph, we already have $S_1$. The other sum is
\[ S_2 = \frac{1}{N} \frac{N-1}{N^2} \approx \frac{1}{N^2}. \]
Thus $S_1 / \sqrt{S_2} \approx 1$, and so the success probability roughly reaches $1$ at time $\pi \sqrt{N} / 2$, as expected from the introduction.


\section{Chiral Quantum Search}

Now we apply the method of Childs and Goldstone, which we summarized in the previous section, to the chiral quantum walk search algorithm. 

\subsection{Eigenvalues of $-L$}

To calculate the sums $S_1$ and $S_2$, we need the eigenvalues of $-L$. Since $-L$ is a circulant matrix, its eigenvalues $\mathcal{E}_j$ can be expressed in terms of roots of unity $\omega_j = e^{2\pi ij/N}$ \cite{Davis1994}:
\[ \mathcal{E}_j = (N-1)\cos\theta - \rme^{-\rmi\theta} \mathrm{\Sigma}_j - \rme^{\rmi\theta} \mathrm{\Sigma}_j^*, \]
where $j = 0, 1, \dots, N-1$,
\[ \mathrm{\Sigma}_j = \omega_j + \omega_j^2 + \dots + \omega_j^{\frac{N-1}{2}} \]
is the sum of the roots of unity in the upper-half complex plane, and
\[ \mathrm{\Sigma}_j^* = \omega_j^{\frac{N-1}{2}+1} + \omega_j^{\frac{N-1}{2}+2} + \dots + \omega_j^{N-1} \]
is its conjugate and corresponds to the sum of the roots of unity in the lower-half complex plane. Note that $\omega_0 = 1$, and so $\mathcal{E}_0 = 0$. Since $\mathrm{\Sigma}_j$ is a geometric series, when $j \ne 0$,
\[ \mathrm{\Sigma}_j = \frac{\omega_j - \omega_j^{\frac{N-1}{2}+1}}{1-\omega_j}. \]
To further simplify this sum, we multiply the top and bottom of it by $1 - \omega_j^*$:
\begin{eqnarray*}
	\mathrm{\Sigma}_j
	&= \frac{\omega_j + \left( \omega_j^{\frac{N-1}{2}} - \omega_j^{\frac{N-1}{2}+1} \right) - 1}{2 - (\omega_j + \omega_j^*)}
	= \frac{\omega_j + 2 \rmi \,{\rm Im}\!\left(\omega_j^{\frac{N-1}{2}}\right) - 1}{2 - 2 \,{\rm Re}(\omega_j)} \\
	&= -\frac{1}{2} + \rmi \frac{\sin\left(\frac{2\pi j}{N}\right) + 2 \sin\left(\frac{2\pi j}{N} \frac{N-1}{2}\right)}{2 - 2\cos\left(\frac{2\pi j}{N}\right)} = -\frac{1}{2} + \rmi\frac{\alpha_j}{2},
\end{eqnarray*}
where we have defined
\[ \alpha_j = \frac{\sin\left(\frac{2\pi j}{N}\right) + 2 \sin\left(\frac{2\pi j}{N} \frac{N-1}{2}\right)}{1 - \cos\left(\frac{2\pi j}{N}\right)}. \]
We will further simplify $\alpha_j$ in a moment, but first let us show how this expression for $\mathrm{\Sigma}_j$ simplifies our expression for the eigenvalues of $-L$ with $j \ne 0$:
\begin{eqnarray}
	\mathcal{E}_{j \ne 0} 
	&= (N-1)\cos\theta - \rme^{-\rmi\theta} \left( -\frac{1}{2} + \rmi\frac{\alpha_j}{2} \right) - \rme^{\rmi\theta} \left( -\frac{1}{2} - \rmi\frac{\alpha_j}{2} \right) \nonumber \\
	&= (N-1)\cos\theta + \frac{1}{2} \left( \rme^{\rmi\theta} + \rme^{-\rmi\theta} \right) + \rmi\frac{\alpha_j}{2} \left( \rme^{\rmi\theta} - \rme^{-\rmi\theta} \right) \nonumber \\
	&= (N-1)\cos\theta + \cos\theta - \alpha_j\sin\theta \nonumber \\
	&= N \cos\theta - \alpha_j\sin\theta \label{eq:E_Laplacian}.
\end{eqnarray}
Thus we have a simple expression for the eigenvalues of $-L$.

Now let us return to simplifying $\alpha_j$. For its first sine term, we use $\sin(2u) = 2\sin(u)\cos(u)$. For its second sine term, note that it equals $\sin(\pi j - \pi j/N)$, so we can use $\sin(u+v) = \sin(u)\cos(v) - \sin(v)\cos(u)$, for which the first piece is zero:
\[ \alpha_j = \frac{2\sin\left(\frac{\pi j}{N}\right)\cos\left(\frac{\pi j}{N}\right) - 2 \sin\left(\frac{\pi j}{N}\right)\cos(\pi j)}{1 - \cos\left(\frac{2\pi j}{N}\right)}. \]
Now we factor $2\sin(\pi j/N)$ in the numerator and use $2 \sin^2(u/2) = 1 - \cos(u)$ in the denominator:
\[ \alpha_j = \frac{2\sin\left(\frac{\pi j}{N}\right) \left[ \cos\left(\frac{\pi j}{N}\right) - \cos(\pi j) \right]}{2 \sin^2 \left( \frac{\pi j}{N} \right)} = \frac{\cos\left(\frac{\pi j}{N}\right) - \cos(\pi j)}{\sin \left( \frac{\pi j}{N} \right)}. \]
Note that $\cos(\pi j) = 1$ when $j$ is even and $-1$ when $j$ is odd. So we write each case separately:
\begin{eqnarray}
	\alpha_{j,{\rm odd}} = \frac{\cos\left(\frac{\pi j}{N}\right) + 1}{\sin \left( \frac{\pi j}{N} \right)} = \cot\left(\frac{\pi j}{N}\right) + \csc\left(\frac{\pi j}{N}\right) \label{eq:alpha_odd} \\
	\alpha_{j,{\rm even}} = \frac{\cos\left(\frac{\pi j}{N}\right) - 1}{\sin \left( \frac{\pi j}{N} \right)} = \cot\left(\frac{\pi j}{N}\right) - \csc\left(\frac{\pi j}{N}\right). \label{eq:alpha_even}
\end{eqnarray}
Plugging these into \eref{eq:E_Laplacian} gives simple expressions for $\mathcal{E}_{j \ne 0}$.

\subsection{Critical Thetas}

\begin{figure}
\begin{center}
	\includegraphics{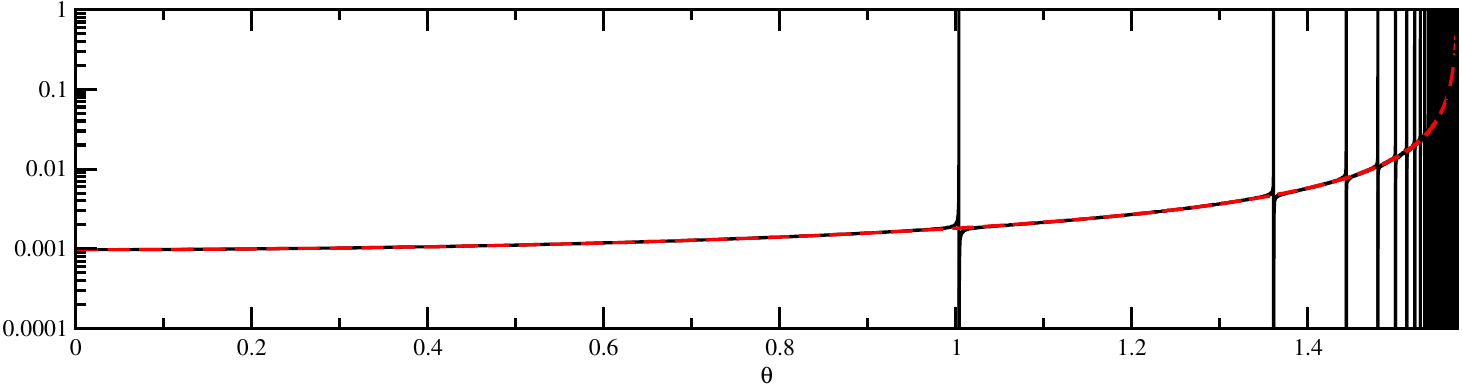}
	\caption{\label{fig:S1_cos_N1023} The sum $S_1$ (solid black) and its approximation $1/(N\cos\theta)$ (dashed red) as a function of $\theta$ with $N = 1023$.}
\end{center}
\end{figure}

Now that we have the eigenvalues $\mathcal{E}_j$ of $-L$, we can find the sum \eref{eq:CG_sum}
\[ S_1 = \frac{1}{N} \sum_{\mathcal{E}_j \ne 0} \frac{1}{\mathcal{E}_j}, \]
which is approximately the critical $\gamma$ for large $N$ \eref{eq:CG_gammac}. Since $\mathcal{E}_0 = 0$, we exclude the $j = 0$ term in the sum. But for certain ``critical'' values of $\theta$, there are other eigenvalues $\mathcal{E}_{j\ne0}$ that are also zero, at which $S_1$ diverges. This is shown in \fref{fig:S1_cos_N1023}; $S_1$ diverges at several critical values of $\theta$, and they are more frequent as $\theta$ approaches $\pi/2$.

We can analytically find the $\theta_c$'s corresponding to these divergences. Since $\alpha_{j,{\rm odd}}$ \eref{eq:alpha_odd} is always positive and $\alpha_{j,{\rm even}}$ \eref{eq:alpha_even} is always negative, $\mathcal{E}_{j\ne0}$ \eref{eq:E_Laplacian} can only be zero when $j$ is odd. Setting $\mathcal{E}_{j,{\rm odd}} = 0$ and solving for $\theta$, we get
\[ \theta_c = \tan^{-1} \left( \frac{N}{\alpha_{j,{\rm odd}}} \right). \]
This can be further simplified when $j$ scales less than $N$ by Taylor expanding $\alpha_{j,{\rm odd}} \approx 2N/(\pi j)$ for large $N$. Then the critical $\theta$'s are
\[ \theta_c \approx \tan^{-1} \left( \frac{\pi j}{2} \right) \approx 1.0039, 1.3617, 1.4442, 1.4801, 1.5002, \dots . \]
Since $j$ is odd in this expression, let us call these values $\theta_{c1}, \theta_{c3}, \theta_{c5}$, and so on. Note that they are independent of $N$, and they correspond to the divergences of $S_1$ in \fref{fig:S1_cos_N1023}.

\subsection{Critical Gamma}

Now let us return to finding the sum $S_1$, which is approximately the critical $\gamma$ for large $N$ \eref{eq:CG_gammac}. We assume that $\theta$ is away from its critical values. Otherwise, $S_1$ would diverge, and $\gamma_c$ would also diverge, which is unphysical since $\gamma$ is the jumping rate of the randomly walking quantum particle. In particular, we prove below for large $N$ that away from the $\theta_c$'s,
\begin{equation}
	\label{eq:S1}
	S_1 \approx \frac{1}{N \cos\theta},
\end{equation}
as shown in \fref{fig:S1_cos_N1023}. Note when $\theta = 0$, this yields $\gamma_c = 1/N$, as expected from the introduction.

To begin the proof of \eref{eq:S1}, first note from \eref{eq:alpha_odd} and \eref{eq:alpha_even} that $\alpha_{N-j} = -\alpha_j$. Then using \eref{eq:E_Laplacian}, we can rewrite the sum as
\[ S_1 = \frac{1}{N} \sum_{j = 1,\dots,\frac{N-1}{2}} \left( \frac{1}{N \cos\theta - \alpha_j \sin\theta} + \frac{1}{N \cos\theta + \alpha_{j} \sin\theta} \right). \]
The summand takes the form
\[ \frac{1}{a+b} + \frac{1}{a-b} = \frac{2a}{a^2-b^2}, \]
so the sum is
\[ S_1 = 2\cos\theta \sum_{j = 1,\dots,\frac{N-1}{2}} \frac{1}{N^2 \cos^2\theta - \alpha_j^2 \sin^2\theta} . \]
Let us consider this for different ranges of $\theta$. When it is less than its first critical value (\textit{i.e.}, $\theta < \theta_{c1} \approx 1.0039$), then the denominator is dominated by $N^2 \cos^2\theta$, yielding
\[ S_1 \approx 2\cos\theta \sum_{j = 1,\dots,\frac{N-1}{2}} \frac{1}{N^2 \cos^2\theta} = \frac{2\cos\theta}{N^2 \cos^2\theta} \frac{N-1}{2} \approx \frac{1}{N\cos\theta}, \]
which is \eref{eq:S1}. Now consider when $\theta$ is between two consecutive critical values $\theta_{cJ}$ and $\theta_{c(J+2)}$ with $J$ odd. Note that $J$ is small compared to $N$, otherwise $\theta$ would approach $\pi/2$ where $S_1$ diverges (see \fref{fig:S1_cos_N1023}). In this case, we split $S_1$ into two sums with odd and even $j$'s,
\begin{eqnarray*}
	S_{1,{\rm odd}} = 2\cos\theta \sum_{j = 1,3,\dots,\frac{N-1}{2}} \frac{1}{N^2 \cos^2\theta - \alpha_j^2 \sin^2\theta} \\
	S_{1,{\rm even}} = 2\cos\theta \sum_{j = 2,4,\dots,\frac{N-2}{2}} \frac{1}{N^2 \cos^2\theta - \alpha_j^2 \sin^2\theta},
\end{eqnarray*}
and show that each contributes $1/(2N\cos\theta)$ to the total sum. For concreteness, we have assumed $(N-1)/2$ to be odd, but our proof still holds when it is even. Beginning with the odd sum, when $j \le J$, the $\alpha_j^2 \sin^2\theta$ term dominates the denominator, while for $j > J$, the $N^2 \cos^2\theta$ term dominates:
\[ S_{1,{\rm odd}} \approx 2\cos\theta \left( \sum_{j = 1,3,\dots,J} \frac{-1}{\alpha_j^2 \sin^2\theta} + \sum_{j = J+2,J+4,\dots,\frac{N-1}{2}} \frac{1}{N^2 \cos^2\theta} \right). \]
We want to show that the second sum in the parenthesis dominates the first. To do this, we bound each of them. The first can be upper bounded by replacing $\theta$ by $\theta_{c1}$:
\begin{eqnarray*}
	\sum_{j = 1,3,\dots,J} \frac{-1}{\alpha_j^2 \sin^2\theta} 
		&< \sum_{j = 1,3,\dots,J} \frac{-1}{\alpha_j^2 \sin^2\theta_{c1}} \approx \sum_{j = 1,3,\dots,J} \frac{-\pi j}{4N^2 \sin^2\theta_{c1}} \\
		&\approx -\frac{\pi^2+4}{16\pi} \frac{(J+1)^2}{N^2},
\end{eqnarray*}
where we used $\alpha_j \approx 2N/(\pi j)$ for small $j$, $\sin\theta_{c1} \approx \pi/\sqrt{\pi^2+4}$, and $1 + 3 + \dots + J = (J+1)^2/4$. The second sum in the parenthesis can be lower bounded by replacing $\theta$ by $\theta_{c1}$:
\begin{eqnarray*}
	\sum_{j = J+2,J+4,\dots,\frac{N-1}{2}} \frac{1}{N^2 \cos^2\theta} 
		&> \sum_{j = J+2,J+4,\dots,\frac{N-1}{2}} \frac{1}{N^2 \cos^2\theta_{c1}} \\
		&\approx \frac{\pi^2+4}{16} \frac{N-2J-1}{N^2},
\end{eqnarray*}
where we used $\cos\theta_{c1} \approx 2/\sqrt{\pi^2+4}$ and that the sum has $(N-2J-1)/4$ terms. From these, we see that the second sum in the parenthesis dominates over the first when $J = o(\sqrt{N})$, which is true so that $\theta$ is not near its critical values. Thus the leading-order behavior of the odd sum is
\begin{eqnarray*}
	S_{1,{\rm odd}} 
		&\approx 2\cos\theta \sum_{j = J+2,J+4,\dots,\frac{N-1}{2}} \frac{1}{N^2 \cos^2\theta} = \frac{2\cos\theta}{N^2\cos^2\theta} \frac{N-2J-1}{4} \\
		&\approx \frac{1}{2N\cos\theta}.
\end{eqnarray*}
Now to complete the proof that $S_1 \approx 1/(N\cos\theta)$, we note that the even sum $S_{1,{\rm even}}$ is always dominated by $N^2 \cos^2\theta$ in its denominator when $\theta$ does not approach $\pi/2$. Then it is also approximately $1/(2N\cos\theta)$ for large $N$, which when combined with the odd sum yields \eref{eq:S1}. $\square$

\begin{figure}
\begin{center}
	\subfloat[] {
		\includegraphics[width=1.8in]{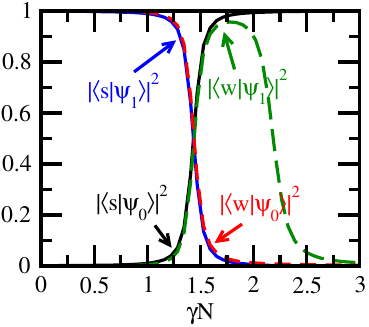}
		\label{fig:overlap_N1023_theta08}
	} \enspace
	\subfloat[] {
		\includegraphics[width=1.8in]{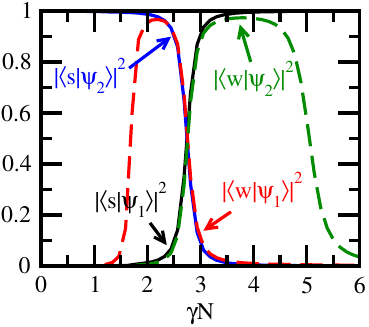}
		\label{fig:overlap_N1023_theta12}
	} \enspace
	\subfloat[] {
		\includegraphics[width=1.8in]{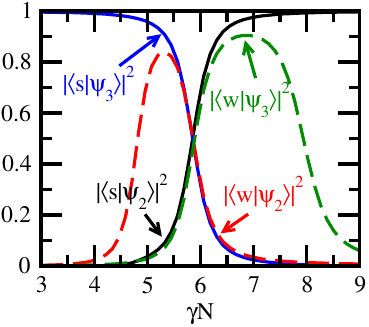}
		\label{fig:overlap_N1023_theta14}
	}
	\caption{\label{fig:overlap_N1023} Probability overlaps of $\ket{s}$ and $\ket{w}$ with the eigenenergies $\ket{\psi_i}$ of $H$ for search by chiral quantum walk with $N = 1023$ and (a) $\theta = 0.8$, (b) $\theta = 1.2$, and (c) $\theta = 1.4$.}
\end{center}
\end{figure}

For nonzero values of $\theta$, some overlap plots similar to \fref{fig:overlap_N1023_theta0} are shown in \fref{fig:overlap_N1023}. Let us discuss each of them. In \fref{fig:overlap_N1023_theta08}, we have $\theta = 0.8$, and our formula \eref{eq:S1} yields the critical $\gamma_c N \approx S_1 N \approx 1.44$. This corresponds to where the ground and first excited states of the Hamiltonian are proportional to $\ket{s} \mp \ket{w}$, so the system rotates from $\ket{s}$ to $\ket{w}$, as desired. In the next plot, \fref{fig:overlap_N1023_theta12} shows a similar behavior when $\theta = 1.2$, but now at $\gamma_c N \approx S_1 N \approx 2.76$ and with the first and second excited states. So we still expect the system to rotate from $\ket{s}$ to $\ket{w}$, but the evolution approximately occurs in the 2D subspace spanned by $\ket{\psi_1}$ and $\ket{\psi_2}$ instead of $\ket{\psi_0}$ and $\ket{\psi_1}$. Similarly in \fref{fig:overlap_N1023_theta14}, for which $\theta = 1.4$, the 2D support has shifted again, this time to $\ket{\psi_2}$ and $\ket{\psi_3}$ at the critical $\gamma_c N \approx S_1 N \approx 5.88$. Next, we will explain why these shifts occur as $\theta$ increases and derive the critical $\theta$'s at which these transitions occur.

\subsection{2D Subspace Support}

\begin{figure}
\begin{center}
	\includegraphics[width=1.8in]{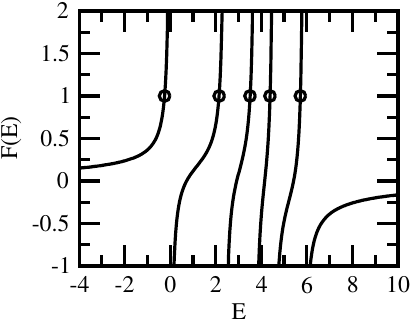}
	\caption{\label{fig:F_N5_theta06_gamma1} $F(E)$ for $N = 5$, $\theta = 0.6$, and $\gamma = 1$. As indicated by circles, $F(E) = 1$ corresponds to eigenenergies of the search Hamiltonian.}
\end{center}
\end{figure}

To explain why the energy levels that support the system change with $\theta$, we must look at some details of Childs and Goldstone's \cite{CG2004} method, which we ignored in our summary in Section~\ref{Section:CG}. From their (26), the eigenvalues $\mathcal{E}_j$ of $-L$ can be used to find the eigenenergies $E_a$ of the search Hamiltonian \eref{eq:H} by solving
\[ F(E_a) = 1, \]
where
\[ F(E) = \frac{1}{N} \sum_j \frac{1}{\gamma \mathcal{E}_j - E}. \]
An example of $F(E)$ is plotted in \fref{fig:F_N5_theta06_gamma1}, and we give some properties of it that were identified by Childs and Goldstone. $F(E)$ has poles where $E = \gamma \mathcal{E}_j$, $F'(E) > 0$ everywhere (except at the poles), and $F(E) \to 0$ as $E \to \pm \infty$. So there exists an eigenvalue $E_a$ of $H$ between every adjacent pair of poles, and one left of the smallest pole. If $\theta$ is sufficiently small so that $\mathcal{E}_0 = 0$ is the lowest eigenvalue of $-L$, then the ground state energy is negative, and the other eigenenergies are positive.

As $\theta$ increases, however, then \eref{eq:E_Laplacian} and \eref{eq:alpha_odd} imply that $\mathcal{E}_j$ decreases when $j$ is odd. In particular, $\mathcal{E}_j = 0$ at the critical $\theta$'s from earlier:
\begin{eqnarray*}
	\theta_c &= \tan^{-1} \left( \frac{N}{\alpha_{j,{\rm odd}}} \right) \\
		&\approx \tan^{-1} \left( \frac{\pi j}{2} \right) \approx 1.0039, 1.3617, 1.4442, 1.4801, 1.5002, \dots,
\end{eqnarray*}
at which a level crossing (degeneracy) occurs in the eigenenergies of $H$. As $\theta$ continues to increase beyond critical values, eigenenergies of $H$ become negative, and the system approximately evolves in higher and higher energy eigenstates of $H$.

\begin{figure}
\begin{center}
	\includegraphics{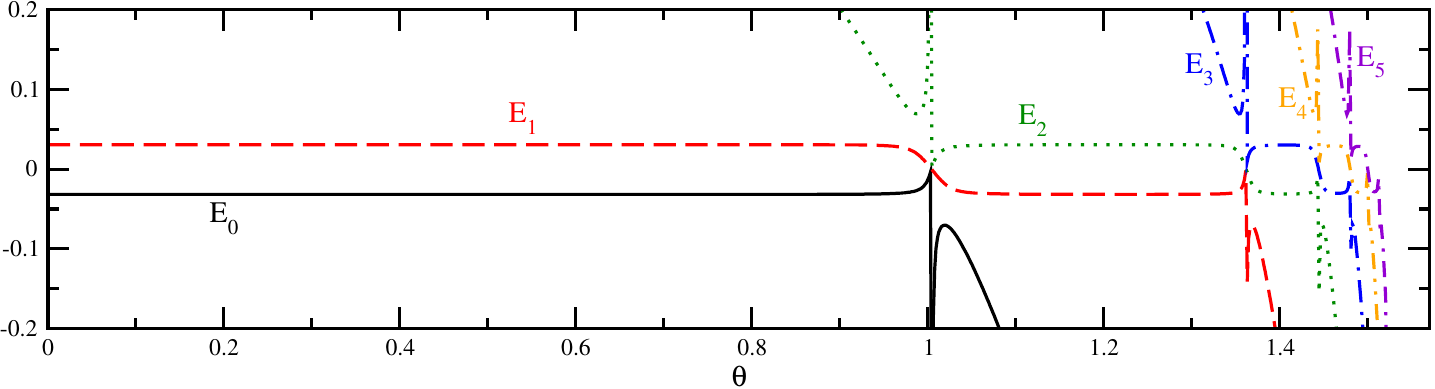}
	\caption{\label{fig:energies_N1023} The lowest six eigenenergies of the chiral quantum walk search Hamiltonian \eref{eq:H} with $N = 1023$ and $\gamma = S_1$.}
\end{center}
\end{figure}

The lowest six eigenenergies of $H$ are shown in \fref{fig:energies_N1023} as $\theta$ is increased. When $\theta$ is sufficiently below the first critical value of $1.0039$, the system remains in its ground and first excited states. Since the energy gap is constant throughout this region, the runtime is unaffected by $\theta$. When $\theta$ crosses the first critical value so that it is sufficiently between $1.0039$ and $1.3617$, the system transitions to being in the first and second excited states, with the same energy gap as before. So the runtime is unchanged, even though higher energy eigenstates of $H$ support the evolution. This continues as each $\theta_c$ is crossed, until the critical values occur so close together that there is no opportunity for the system to settle into a 2D subspace. Note that the sharp behavior near each $\theta_c$ is because we used $\gamma = S_1$, which diverges at the critical $\theta$'s (recall \fref{fig:S1_cos_N1023}).

Childs and Goldstone's method follows the two eigenenergies closest to 0, which is evident from the derivation of their method \cite{CG2004}. So as the eigenenergies become more and more negative, they ``drop out,'' and the method continues to track the two correct energy eigenstates.

\subsection{Success Probability and Runtime}

\begin{figure}
\begin{center}
	\includegraphics{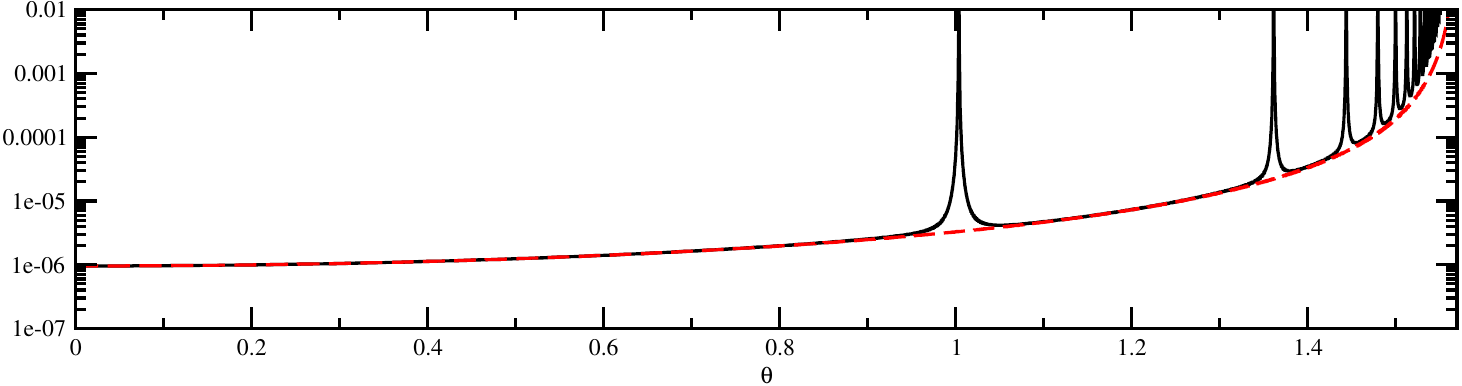}
	\caption{\label{fig:S2_cos_N1023} The sum $S_2$ (solid black) and its approximation $1/(N^2\cos^2\theta)$ (dashed red) as a function of $\theta$ with $N = 1023$.}
\end{center}
\end{figure}

From \fref{fig:energies_N1023} and its description in the previous section, we see that the energy gap between the two relevant eigenstates is constant away from the critical $\theta$'s. So the success probability should reach $1$ at time $\pi \sqrt{N} / 2$. To prove this analytically, we use \eref{eq:CG_prob} and \eref{eq:CG_runtime}, for which we need the sum \eref{eq:CG_sum}
\[ S_2 = \frac{1}{N} \sum_{\mathcal{E}_j \ne 0} \frac{1}{\mathcal{E}_j^2}. \]
Finding the behavior of this for $\theta$ away from its critical values follows the same procedure as our proof of $S_1$ \eref{eq:S1}, yielding
\begin{equation}
	\label{eq:S2}
	S_2 \approx \frac{1}{N^2 \cos^2\theta}.
\end{equation}
A plot of $S_2$ with this approximation is shown in \fref{fig:S2_cos_N1023}. Note when $\theta = 0$, this yields $S_2 = 1/N^2$, as expected from the introduction.

Now that we have expressions for both $S_1$ \eref{eq:S1} and $S_2$ \eref{eq:S2}, the success probability \eref{eq:CG_prob} of the chiral quantum walk search algorithm reaches
\[ p_* \approx \frac{S_1}{\sqrt{S_2}} \approx \frac{\frac{1}{N\cos\theta}}{\sqrt{\frac{1}{N^2\cos^2\theta}}} = 1, \]
at time \eref{eq:CG_runtime}
\[ t_* \approx \frac{\pi}{2} \frac{\sqrt{S_2 N}}{S_1} \approx \frac{\pi}{2} \sqrt{N}. \]
So even with time-reversal symmetry broken, the runtime and success probability is unchanged, so long as we are away from the critical $\theta$'s at which the eigenvectors of $H$ that support the evolution change.

\begin{figure}
\begin{center}
	\subfloat[] {
		\includegraphics[width=1.8in]{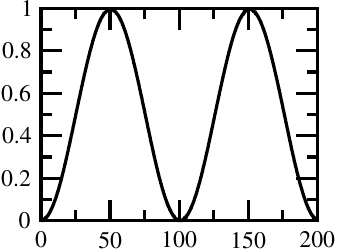}
		\label{fig:prob_time_N1023_theta08}
	} \quad
	\subfloat[] {
		\includegraphics[width=1.8in]{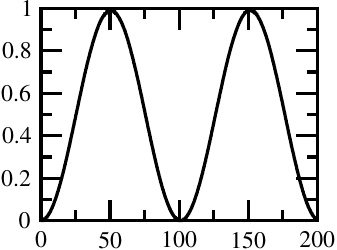}
		\label{fig:prob_time_N1023_theta12}
	} \quad
	\subfloat[] {
		\includegraphics[width=1.8in]{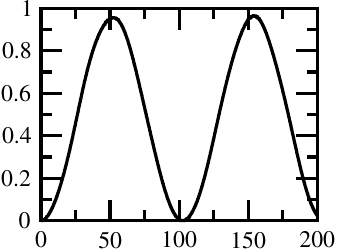}
		\label{fig:prob_time_N1023_theta14}
	}
	\caption{\label{fig:prob_time_N1023} Success probability as a function of time for chiral quantum walk search with $N = 1023$, $\gamma = S_1$, and (a) $\theta = 0.8$, (b) $1.2$, and (c) $1.4$.}
\end{center}
\end{figure}

These success probabilities and runtimes are confirmed in \fref{fig:prob_time_N1023}, where we plot the evolution of the success probability using the same values of $\theta$ that we used in \fref{fig:overlap_N1023}. The success probabilities roughly reach $1$ at time $\pi \sqrt{1023} / 2 \approx 50.241$. The last figure with $\theta = 1.4$ is a little short because our approximations for $S_1$ and $S_2$ are less accurate as each critical $\theta$ is passed, so bigger $N$ is needed for the probability to near $1$.

\section{Conclusion}

Time-reversal symmetry breaking in quantum walks was first exploited for quantum transport, yielding significant speedups. Here, however, we considered the first algorithmic investigation of such chiral quantum walks, defining an unstructured search problem on a modified complete graph that retains the homogeneity of the search problem.  While breaking time-reversal symmetry does affect the jumping rate $\gamma$ that must be used, and can also change the energy eigenstates in which the system evolves, it does not change the runtime of the search for large swathes of $\theta$. Thus quantum search is robust against time-reversal symmetry breaking in the model we have introduced.


\ack
Thanks to Jacob Biamonte for useful discussions. This work was supported by the European Union Seventh Framework Programme (FP7/2007-2013) under the QALGO (Grant Agreement No.~600700) project, and the ERC Advanced Grant MQC.


\section*{References}
\bibliographystyle{iopart-num}
\bibliography{refs}

\end{document}